\documentclass[conference]{IEEEtran}
\IEEEoverridecommandlockouts

\usepackage[table,xcdraw]{xcolor}
\usepackage{graphicx}
\usepackage{amsmath}
\usepackage{amssymb}
\usepackage{multirow}
\usepackage{xcolor}
\usepackage{colortbl}
\usepackage{pgfplots}
\usepackage{layout}
\usepackage{amssymb}
\usepackage{bbding}
\usepackage{booktabs}
\IfFileExists{orcidlink.sty}{\usepackage{orcidlink}}{\newcommand{\orcidlink}[1]{}}
\usepackage{diagbox}

\usepackage{stfloats}
\usepackage[misc]{ifsym}
\usepackage{makecell}
\usepackage{caption}
\usepackage[normalem]{ulem}

\usepackage[T1]{fontenc}

\usepackage{graphicx,verbatim}
\usepackage{fancyhdr}
\usepackage{hyperref}

\usepackage{booktabs}
\usepackage[table]{xcolor}
\usepackage{amssymb}
\usepackage{pifont}
\usepackage{graphicx}
\usepackage{booktabs}

\usepackage{graphicx}
\usepackage{float}
\usepackage[section]{placeins}

\def\BibTeX{{\rm B\kern-.05em{\sc i\kern-.025em b}\kern-.08em
    T\kern-.1667em\lower.7ex\hbox{E}\kern-.125emX}}

\begin{document}

\title{
CHM-Net: Center Heatmap-driven Macro-Micro Modeling Network for MRI-based Microbial Density Stratification
}


\author{
\centering
\small
\begin{tabular}{@{}c@{\hspace{0.04\textwidth}}c@{}}
\parbox{0.46\textwidth}{\centering 1\textsuperscript{st} Jiaming Liang$^\dagger$\,\orcidlink{0009-0000-6007-9759}\\
\textit{School of Computer Science and Engineering}\\
\textit{South China University of Technology}\\
Guangzhou, China\\
csliangjm@mail.scut.edu.cn} & \parbox{0.46\textwidth}{\centering 2\textsuperscript{nd} Haolin Chen$^\dagger$\,\orcidlink{0009-0001-1831-5276}\\
\textit{School of Future Technology}\\
\textit{South China University of Technology}\\
Guangzhou, China\\
hlchen@mail.scut.edu.cn}\\[1.0em]
\parbox{0.46\textwidth}{\centering 3\textsuperscript{rd} Tingting Li\\
\textit{Department of Medical Imaging}\\
\textit{Affiliated Cancer Hospital, Guangzhou Medical University}\\
Guangzhou, China\\
18296617132@163.com} & \parbox{0.46\textwidth}{\centering 4\textsuperscript{th} Bowen Yu\,\orcidlink{0009-0001-8792-9044}\\
\textit{School of Computer Science and Engineering}\\
\textit{South China University of Technology}\\
Guangzhou, China\\
im.yubw@gmail.com}\\[1.0em]
\parbox{0.46\textwidth}{\centering 5\textsuperscript{th} Qianyan Long\\
\textit{Department of Medical Imaging}\\
\textit{Affiliated Cancer Hospital, Guangzhou Medical University}\\
Guangzhou, China\\
qianyanlongqq@163.com} & \parbox{0.46\textwidth}{\centering 6\textsuperscript{th} Tinghe Zhang\,\orcidlink{0000-0002-4595-6167}\\
\textit{School of Future Technology}\\
\textit{South China University of Technology}\\
Guangzhou, China\\
zhangth@scut.edu.cn}\\[1.0em]
\parbox{0.46\textwidth}{\centering 7\textsuperscript{th} Xi Zhong\,\orcidlink{0000-0002-0713-105X}\\
\textit{Department of Medical Imaging}\\
\textit{Affiliated Cancer Hospital, Guangzhou Medical University}\\
Guangzhou, China\\
zhongxi@gzhmu.edu.cn} & \parbox{0.46\textwidth}{\centering 8\textsuperscript{th} Xiaowei Hu$^*$\,\orcidlink{0000-0002-5708-7018}\\
\textit{School of Future Technology}\\
\textit{South China University of Technology}\\
Guangzhou, China\\
huxiaowei@scut.edu.cn}\\[1.0em]
\parbox{0.46\textwidth}{\centering 9\textsuperscript{th} Xiaoqi Sheng$^*$\,\orcidlink{0000-0002-2929-5805}\\
\textit{School of Future Technology}\\
\textit{South China University of Technology}\\
Guangzhou, China\\
xqsheng@scut.edu.cn} & \parbox{0.46\textwidth}{\centering 10\textsuperscript{th} Hongmin Cai\,\orcidlink{0000-0002-2747-7234}\\
\textit{School of Future Technology}\\
\textit{South China University of Technology}\\
Guangzhou, China\\
hmcai@scut.edu.cn}
\end{tabular}
}
\maketitle
\fancypagestyle{firstpagefooter}{%
  \fancyhf{}%
  \renewcommand{\headrulewidth}{0pt}%
  \renewcommand{\footrulewidth}{0pt}%
  \fancyfoot[L]{\begin{minipage}{0.94\columnwidth}\footnotesize
  $^\dagger$ These authors contributed equally to this work.\\
  $^*$ Corresponding authors: Xiaowei Hu (huxiaowei@scut.edu.cn) and Xiaoqi Sheng (xqsheng@scut.edu.cn).
  \end{minipage}}%
}
\thispagestyle{firstpagefooter}
\pagestyle{plain}

\begin{abstract}

Microbial density is clinically important for tumor assessment and treatment decision-making, and recent advances in deep learning suggest that it can be non-invasively inferred from multimodal MRI. In this work, MRI-based Microbial Density Stratification (MRI-MDS) is first investigated as a patient-level representation learning task, and Center Heatmap-driven Macro-micro modeling Network (CHM-Net) is introduced for this task. CHM-Net first establishes the link between imaging phenotypes and microbial states through center heatmap-guided small-lesion response localization. Building upon this, it constructs patient-level macro-micro evidence from localized heatmap responses for microbial density prediction. Experiments on the novel GBNPC 2026 dataset constructed for MRI-MDS demonstrate the effectiveness of CHM-Net, achieving superior performance over representative baselines with a 12.06\% absolute ACC gain over the strongest competing result. Additionally, auxiliary validation on two 3D medical image datasets further verifies its robustness across volumetric medical image classification scenarios. The project is available at \href{https://anonymous.4open.science/r/CHM-Net-942E/}{URL}.

\end{abstract}


\section{Introduction}

Microbial density is increasingly recognized as an informative indicator of tumor ecology and immune activity~\cite{gopalakrishnan2018gut,nejman2020human,sepichpoore2021microbiome}, with its potential value for tumor microenvironment assessment. However, current microbial density assessment usually requires pathological tissue sections obtained after surgery, making the process invasive and unable to provide microbial density information before treatment. Therefore, an efficient and non-invasive strategy for estimating microbial density from preoperative imaging is clinically expected. From a practical standpoint, existing studies provide biological and imaging evidence for this idea. Specifically, Riquelme et al.~\cite{riquelme2019tumor} find that tumor microbiome composition and diversity are closely associated with the local immune microenvironment, while Aerts et al.~\cite{aerts2014decoding} and Gillies et al.~\cite{gillies2016radiomics} demonstrate that non-invasive imaging features can decode tumor phenotypes and provide quantitative phenotype information. Furthermore, Xiao et al.~\cite{xiao2021dynamic} show that multidimensional dynamic contrast-enhanced MRI features are associated with pathological angiogenesis and microvessel density, indicating that MRI can reflect tumor microenvironmental phenotypes. Therefore, microbial-density-related tissue structure and spatial heterogeneity may be manifested as macroscopic imaging phenotypes in multimodal MRI although routine MRI cannot directly observe microorganisms. Based on this understanding, MRI-based microbial density stratification (MRI-MDS) can be formulated as a patient-level classification task that predicts postoperative microbial density status from preoperative multimodal MRI, thereby motivating deep learning-based non-invasive analysis for MRI-MDS.

However, the subtle and spatially dispersed evidence in MRI-MDS distinguishes it from conventional classification tasks, making direct transfer of existing classification models non-trivial. Specifically, 2D slice-based~\cite{huang2023missformer} or CNN-based\cite{simonyan2015very,huang2017densely,tan2019efficientnet} methods may be dominated by global anatomical or background patterns, while Transformer-based\cite{liu2022video,manzari2023medvit,shamshad2023transformers} or SSM-based\cite{zheng2025xfmamba,liang2025hwaunetr,gu2024mamba} models may still struggle to focus on small lesions or weak-response regions with limited samples. Therefore, MRI-MDS calls for \textbf{response-aware macro-micro evidence modeling} to discover subtle local evidence from multimodal MRI volumes. Specifically, this task involves four key challenges: (1) \textbf{Small-lesion Response Localization}. Microbial-density-related evidence in MRI-MDS may appear as small lesions, weak-response regions, or subtle cross-modal variations, making reliable response localization necessary. (2) \textbf{Local-to-Prediction Association}. Local imaging cues related to microbial density are often subtle and embedded in large background regions. Even when potential high-response areas can be identified, they may not directly provide reliable evidence for patient-level prediction. (3) \textbf{Patient-level Macro-Micro Fusion}. Effective evidence may be distributed across multiple local regions, requiring the model to fuse local micro-level evidence with global MRI semantics.

\begin{figure}[t]
\centering
\setlength{\tabcolsep}{1pt}

\begin{minipage}{0.32\linewidth}
    \centering
    \includegraphics[width=\linewidth]{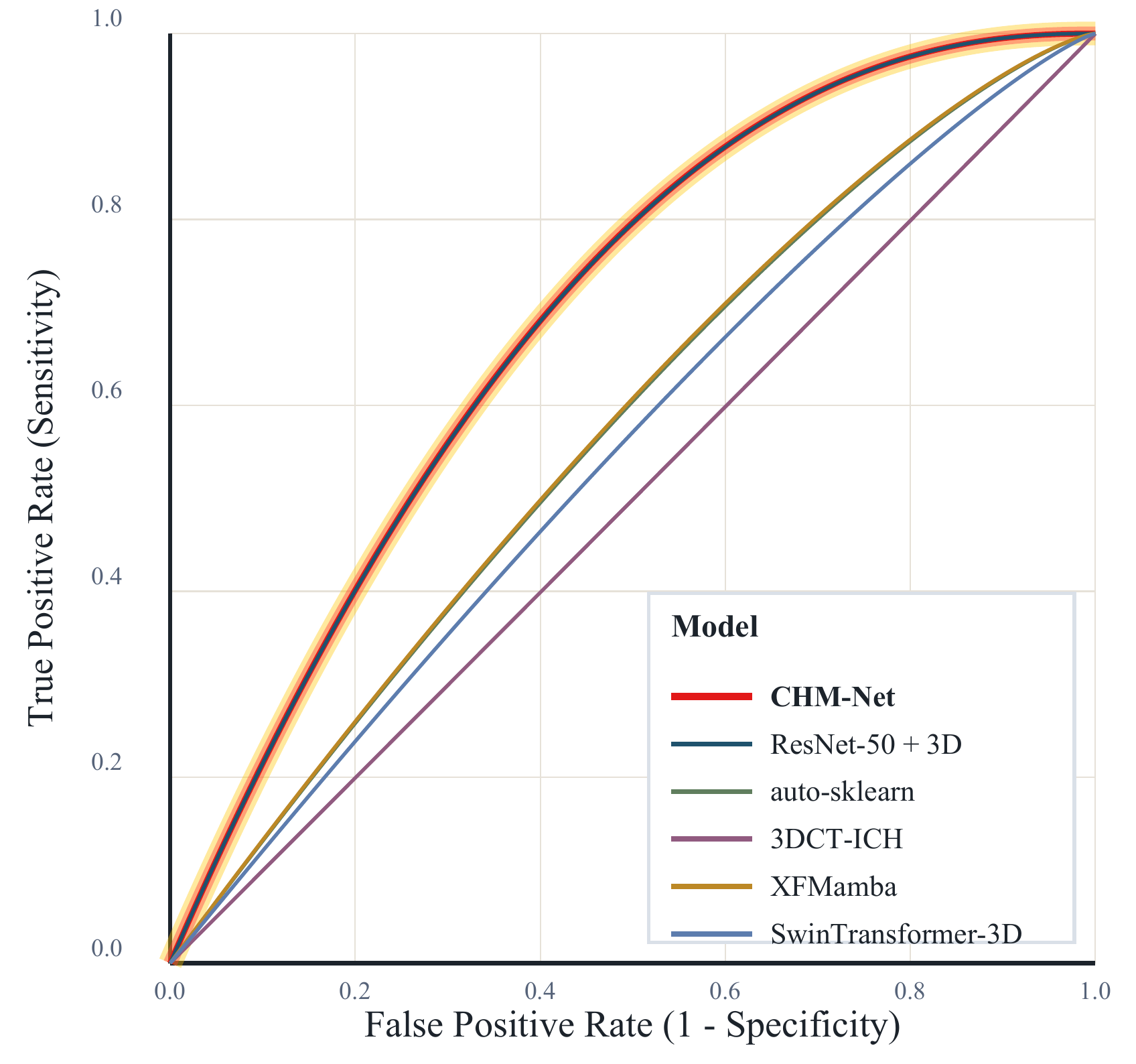}
    \vspace{-1mm}
    \centerline{\scriptsize (a) GBNPC 2026}
\end{minipage}
\hfill
\begin{minipage}{0.32\linewidth}
    \centering
    \includegraphics[width=\linewidth]{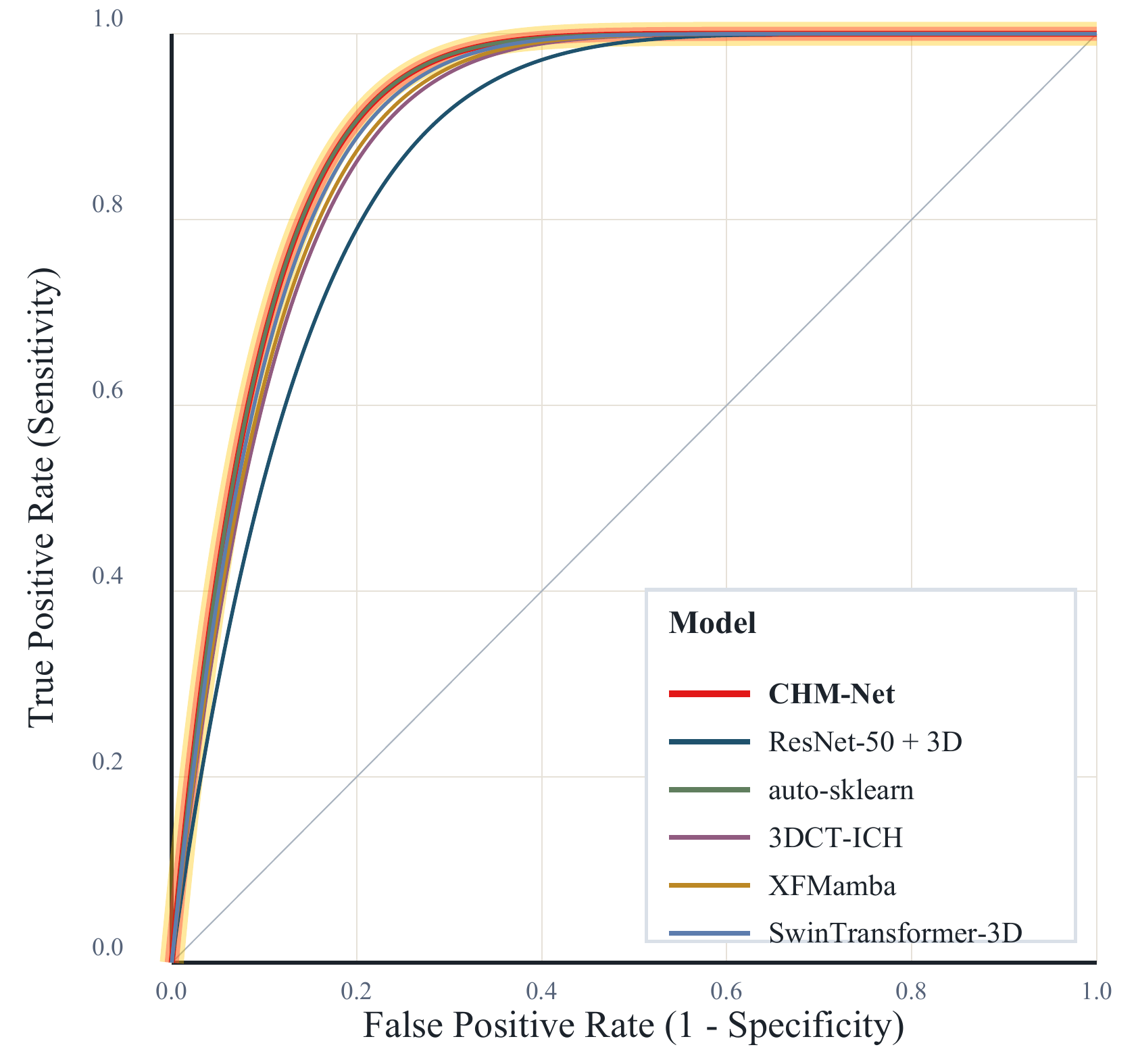}
    \vspace{-1mm}
    \centerline{\scriptsize (b) NoduleMNIST3D}
\end{minipage}
\hfill
\begin{minipage}{0.32\linewidth}
    \centering
    \includegraphics[width=\linewidth]{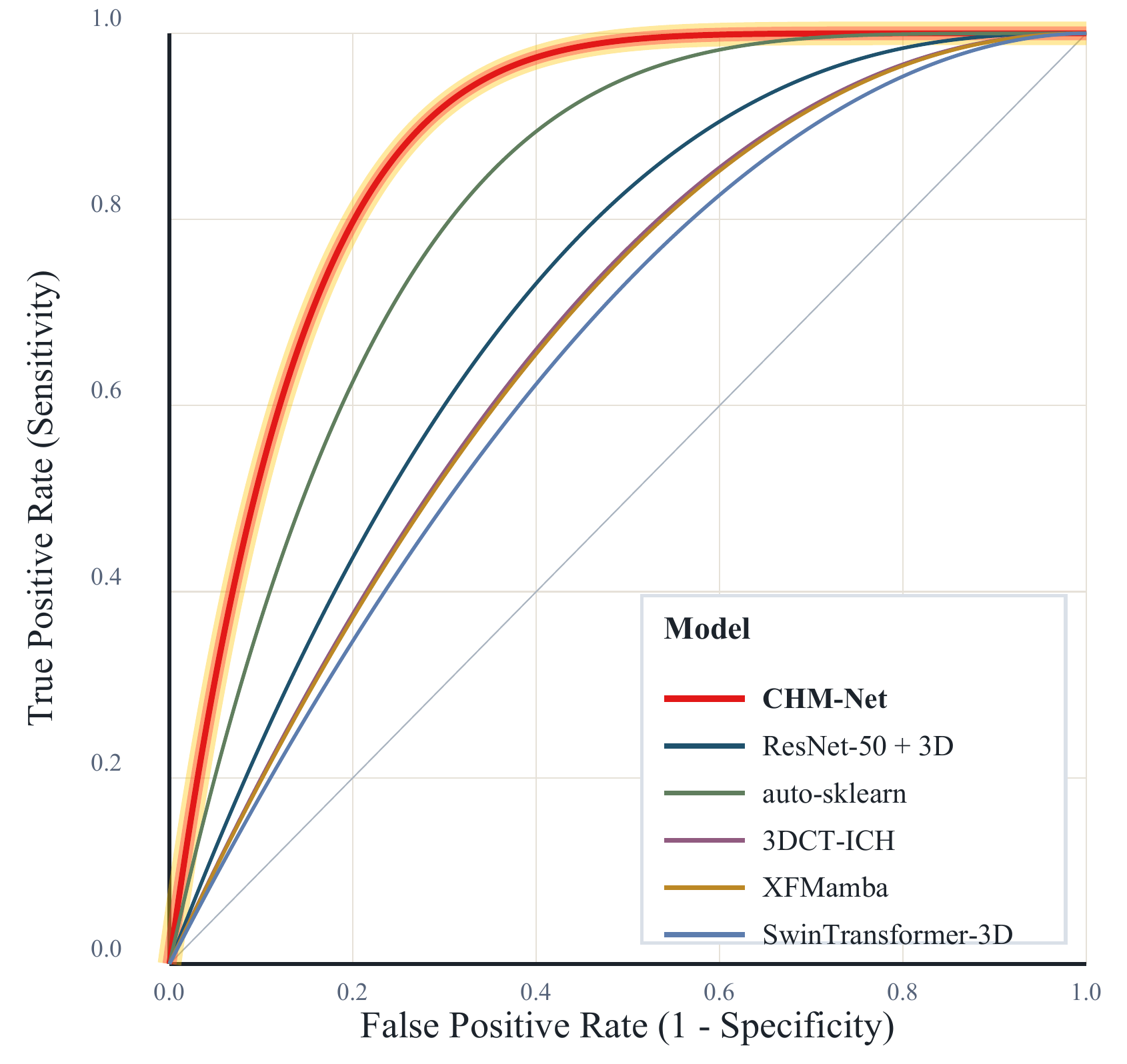}
    \vspace{-1mm}
    \centerline{\scriptsize (c) AdrenalMNIST3D}
\end{minipage}

\caption{ROC-AUC curves on validation datasets}
\label{fig:intro_auc_results}
\end{figure}

To address these challenges and establish a consistent mapping from macroscopic imaging phenotypes to microscopic density states, this work proposes Center Heatmap-driven Macro-micro modeling Network (CHM-Net) as a baseline framework for MRI-MDS. CHM-Net consists of four modules, including Macro Response Transformer (MRT) module, Lesion-aware ROI Miner (LRM) , Tri-planar Micro Encoder (TME) , and Macro-Micro MIL Fusion (MMF) module. Specifically, MRT module enables unsupervised center-response modeling for global spatial prior generation, while LRM conducts heatmap-guided ROI evidence construction. TME and MMF module further provide fine-grained local evidence encoding and patient-level macro-micro fusion, thereby supporting robust density stratification. As shown in Fig.~\ref{fig:intro_auc_results}, CHM-Net achieves superior ROC-AUC performance on GBNPC 2026, demonstrating its effectiveness for MRI-MDS, while consistent results on NoduleMNIST3D and AdrenalMNIST3D further indicate its robustness in volumetric medical image classification. Our contributions can be summarized as follows:

\begin{itemize}
    \item \underline{\textbf{\textit{Innovation.}}}
    This study formulates MRI-MDS as a novel patient-level microbial density stratification task from multimodal MRI. Additionally, CHM-Net is introduced as a task-oriented baseline model, and the novel dataset GBNPC 2026 is constructed as the first multimodal MRI dataset for this problem in nasopharyngeal carcinoma.

    \item \underline{\textbf{\textit{Framework.}}}
    This study proposes CHM-Net as a center heatmap-driven macro-micro modeling network that converts global MRI responses into local ROI evidence and adaptively integrates them with patient-level semantics.

    \item \underline{\textbf{\textit{Validation.}}}
    Experiments demonstrate that CHM-Net outperforms representative baselines on GBNPC 2026 by 12.06\% absolute ACC and remains robust on two public 3D medical classification benchmarks.
\end{itemize}

\section{Related Works}

MRI-MDS can be formulated as a discriminative representation learning problem under a classification paradigm. Early approaches mainly relied on 2D slices or local image regions, whose restricted representational capacity limited full exploitation of volumetric spatial context. With the development of deep learning, CNN-based methods have substantially advanced medical image classification~\cite{ronneberger2015unet,cicek20163d,dou20163d,huang2017densely,tan2019efficientnet,yang2021reinventing}. To adapt convolutional representations to volumetric images, ResNet variants~\cite{hara2018can}, Med3D~\cite{chen2019med3d}, and X3D~\cite{feichtenhofer2020x3d} extend CNN modeling to 3D spatial contexts. Furthermore, AMSNet~\cite{wu2022attention} enhances multi-scale feature aggregation, while 3DCT-ICH~\cite{xiong2024multimodality} incorporates multimodal volumetric information. More recently, Transformer-based architectures have been introduced to medical image analysis~\cite{dosovitskiy2021image,hatamizadeh2022unetr,shamshad2023transformers}. SwinTransformer-3D~\cite{liu2021swin,liu2022video} extends window-based self-attention to 3D representation learning, MedViT-3D~\cite{manzari2023medvit} combines convolutional inductive bias with Transformer modeling, and M3T~\cite{jang2022m3t} integrates multi-view representations for volumetric classification. In parallel, recent SSM-based methods explore efficient sequence modeling for volumetric medical images~\cite{xing2024segmamba,pang2024slimunetr,liang2025hwaunetr,gu2024mamba,liu2024vmamba}, while XFMamba~\cite{zheng2025xfmamba} further emphasizes long-range dependency modeling and cross-view feature fusion. Despite these advances, MRI-MDS remains challenging due to subtle local evidence and the semantic gap between macroscopic MRI phenotypes and microscopic microbial states. To address these limitations, this study proposes CHM-Net to integrate global MRI semantics with local evidence for non-invasive microbial density stratification.

\section{Methodology}

\subsection{Preliminaries: Microbial Density Stratification}

The ground-truth labels for MRI-MDS are derived from pathological slides and clinical records. With the introduction of deep learning, imaging representations associated with microbial density status can be automatically captured from multimodal MRI, making non-invasive MRI-MDS feasible. Conceptually, given a multimodal 3D MRI volume $x \in \mathbb{R}^{C \times D \times H \times W}$, where $C$ denotes the number of MRI modalities and $D$, $H$, and $W$ denote the depth, height, and \begin{figure*}[t]
    \centering
    \includegraphics[width=\textwidth]{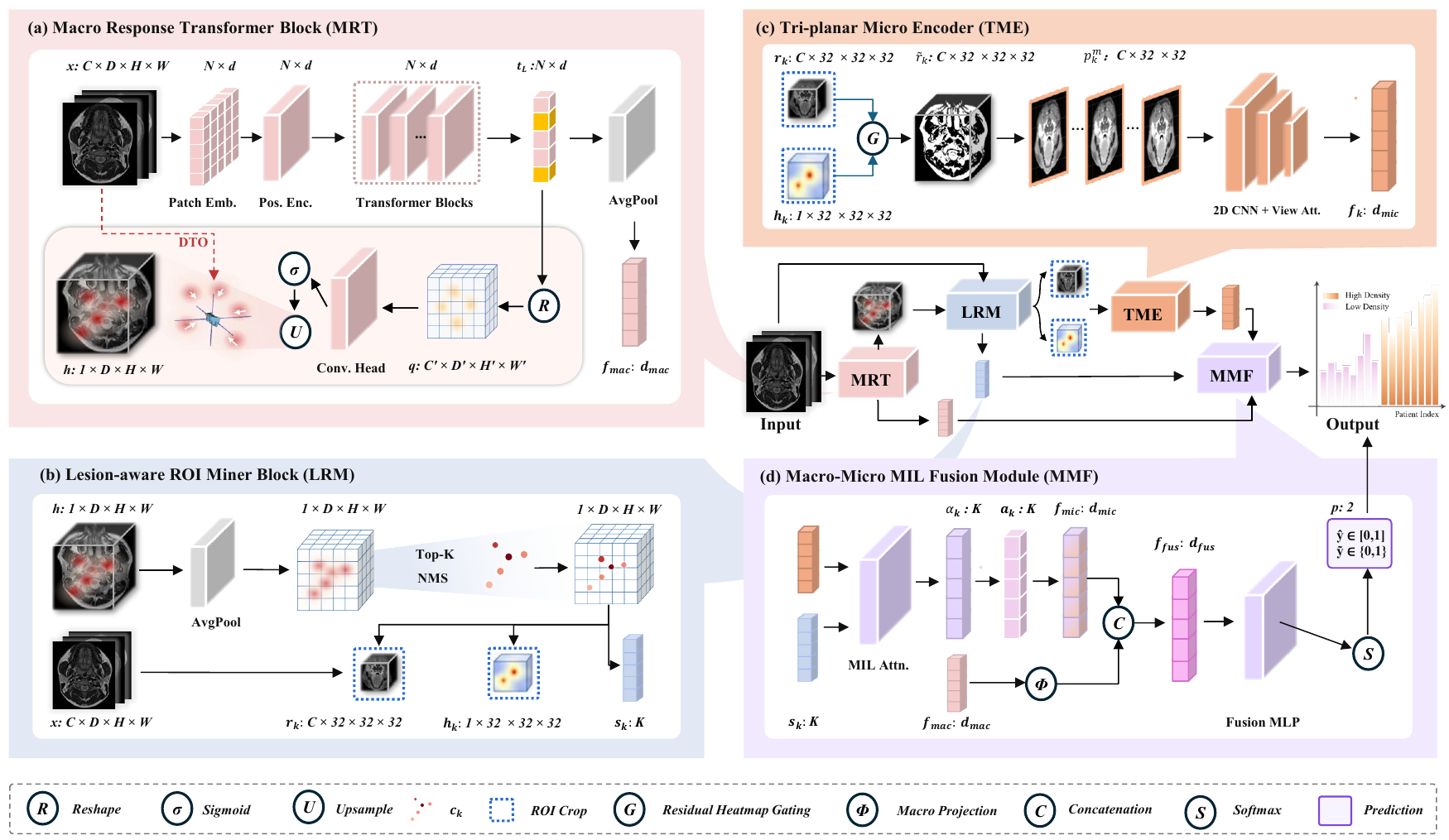}
    \caption{Overall architecture of the proposed CHM-Net.}
    \label{fig:workflow}
\end{figure*}width, respectively, the model outputs a Softmax-activated class probability vector $\mathbf{p}=[p_0,p_1]$, where the high-density probability $\hat{y}=p_1$ is used as prediction confidence, with $\hat{y}\in[0,1]$. The ground-truth label is denoted by $y \in \{0,1\}$, where $0$ and $1$ indicate low and high density, respectively. The discrete prediction label is denoted by $\tilde{y}$.

\subsection{Overview of CHM-Net}

A framework overview is provided first. \textbf{CHM-Net (Center Heatmap-driven Macro-Micro Modeling Network)} is designed for the emerging MRI-MDS task. To address the challenges of localizing small-lesion and weak-response regions, transforming center heatmaps into informative ROI candidates, encoding subtle micro-level ROI patterns, and adaptively integrating local evidence with patient-level MRI semantics, CHM-Net consists of four crucial modules: the \textbf{Macro Response Transformer (MRT) module}, the \textbf{Lesion-aware ROI Miner (LRM) }, the \textbf{Tri-planar Micro Encoder (TME)}, and the \textbf{Macro-Micro MIL Fusion (MMF) module}, as illustrated in Fig.~\ref{fig:workflow}.

Specifically, MRT module first models the global macro-level density context from $x$ and generates a center heatmap $h$ as a spatial prior. LRM then selects $K$ high-response centers from $h$ and crops the corresponding local ROIs. TME further encodes each 3D ROI through axial, coronal, and sagittal 2D views to obtain micro-level features. Finally, MMF module treats multiple ROI features as a patient-level bag, aggregates local features using MIL attention, fuses them with the macro feature, and outputs the patient-level high-density confidence $\hat{y}$ and the discrete prediction label $\tilde{y}$.

\subsection{Macro Response Transformer Module}

Since MRI-MDS depends on small lesion regions in macro-level MRI that often require manual annotation, directly learning the association between
macro-level imaging and microbial expression remains
challenging. To address this issue, the \textbf{Macro Response Transformer (MRT) module} is designed to learn macro-level density context in an unsupervised manner and generate a center heatmap prior, as illustrated in Fig.~\ref{fig:workflow}(a). The detailed unsupervised optimization procedure is introduced in Section~\ref{sec:mdt_pretraining}. At the data-processing level, the input $x$ is first partitioned into a token sequence through 3D patch embedding, and positional encoding is added to retain spatial location information. The token sequence is then processed by Transformer blocks to capture long-range dependencies across spatial regions:
\begin{equation}
t_L = T(E_{\mathrm{pat}}(x)+e_{\mathrm{pos}}),
\end{equation}
\noindent where $t_L$ denotes the final token sequence, $T(\cdot)$ denotes the $L$-block Transformer encoder, $E_{\mathrm{pat}}(\cdot)$ denotes 3D patch embedding, and $e_{\mathrm{pos}}$ denotes positional encoding. \begin{figure}[!htbp]
    \centering
    \includegraphics[width=0.88\linewidth]{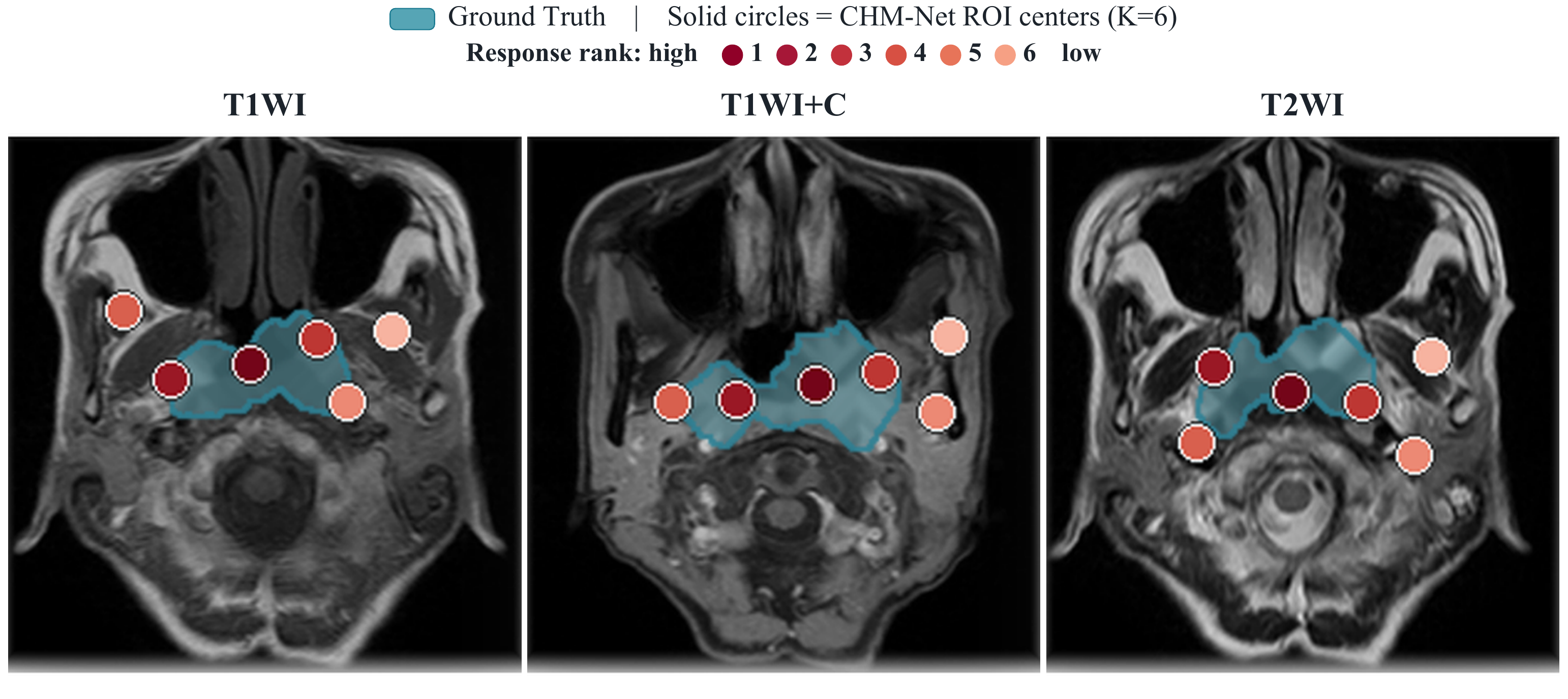}
    \caption{Top-6 ROI Responses Localization.}
    \label{fig:Top-6 ROI Responses Localization}
\end{figure}The encoded tokens are then reshaped into a 3D feature map and used to generate the center heatmap and the macro feature:
\begin{equation}
\begin{gathered}
q = R(t_L), \quad
h = U\big(\sigma(\operatorname{Conv}(q))\big), \\
f_{\mathrm{mac}} = \frac{1}{N}\sum_{i=1}^{N} t_L^i .
\end{gathered}
\end{equation}
\noindent where $q \in \mathbb{R}^{C \times D \times H \times W}$ is the 3D feature map reshaped from the token sequence $t_L$, the center heatmap $h \in [0,1]^{1 \times D \times H \times W}$ serves as a soft spatial prior for subsequent ROI mining, while $f_{\mathrm{mac}}$ preserves patient-level density semantics from the complete MRI volume. $R(\cdot)$ denotes reshaping from tokens to a 3D feature map, $U(\cdot)$ denotes trilinear upsampling, $\sigma(\cdot)$ denotes the sigmoid function, $\operatorname{Conv}(\cdot)$ denotes the convolutional detection head, and $N$ denotes the number of tokens. Thus, MRT module preserves macro-level density information without manual lesion delineation and provides guidance for local ROI mining.

\subsection{Lesion-aware ROI Miner }

Based on the center heatmap generated by MRT module, the \textbf{Lesion-aware ROI Miner (LRM) } converts macro-level spatial prior into local candidate regions for subsequent encoding, aiming to reduce background interference and focus the micro encoder on high-response regions. The procedure of LRM is shown in Fig.~\ref{fig:workflow}(b). Specifically, LRM first applies 3D average pooling to the center heatmap $h$ to suppress isolated noise. The $k$-th ROI center is then selected as the maximum-response position in the currently unsuppressed space:
\begin{equation}
c_k = \arg\max_{v \in \Omega \setminus \mathcal{M}_{k-1}} \operatorname{AvgPool}(h)(v),
\end{equation}
\noindent where $c_k$ denotes the center of the $k$-th ROI, $v$ denotes a candidate voxel coordinate, $\Omega$ denotes the set of all voxel coordinates, and $\mathcal{M}_{k-1}$ denotes the regions suppressed after the previous $k-1$ selections. The input volume $x$ and center heatmap $h$ are cropped around $c_k$:\begin{equation}
r_k = G(x,c_k,S), \quad
h_k = G(h,c_k,S), \quad
s_k = h(c_k),
\end{equation}
\noindent where $r_k \in \mathbb{R}^{C \times S \times S \times S}$ denotes the $k$-th ROI image patch, $h_k \in \mathbb{R}^{1 \times S \times S \times S}$ denotes the local heatmap cropped from $h$, and $s_k$ denotes the ROI-center heatmap response. $G(\cdot)$ denotes 3D cropping, and $S$ denotes the ROI side length, with $S=32$ by default. After each ROI selection, its surrounding region is suppressed within a radius $\rho=16$ by default to avoid repeated sampling of the same high-response location. The number of ROIs is set to $K=6$ by default to balance local feature coverage and computational complexity. As shown in Fig.~\ref{fig:Top-6 ROI Responses Localization}, LRM transforms the macro-level heatmap prior into complementary local regions for subsequent micro-feature extraction after this process.

\subsection{Tri-planar Micro Encoder }


ROI mining provides candidate regions for local feature analysis, while MRI-MDS still requires fine-grained encoding within each ROI. To this end, the \textbf{Tri-planar Micro Encoder (TME) } is introduced to extract micro-level features from the selected ROIs, as illustrated in Fig.~\ref{fig:workflow}(c). For the $k$-th ROI $r_k$ and its local heatmap $h_k$, TME first applies residual heatmap gating to enhance high-response regions while preserving the original ROI information:
\begin{equation}
\tilde{r}_k = r_k \odot (1+\alpha h_k),
\end{equation}
\noindent where $\tilde{r}_k$ denotes the enhanced ROI, $\odot$ denotes element-wise multiplication, and $\alpha$ denotes the heatmap gating strength, with $\alpha=1.0$ by default.

However, directly processing 3D ROIs may increase computational cost and overfitting risk under limited samples, while a single 2D projection may compress spatial information. To address this trade-off, TME generates heatmap-weighted projections from three orthogonal views and extracts ROI-level micro features through a shared 2D CNN and view attention:
\begin{equation}
p_k^m =
\frac{A_m(\tilde{r}_k \odot h_k)}
{A_m(h_k)+\epsilon}, \quad m \in \mathcal{V},
\end{equation}
\begin{equation}
f_k = \sum_{m \in \mathcal{V}} \beta_k^m E_{\mathrm{cnn}}(p_k^m).
\end{equation}
\noindent where $p_k^m$ denotes the heatmap-weighted projection of the $k$-th ROI under view $m$, and $f_k$ denotes the micro feature of the $k$-th ROI. $\mathcal{V}=\{\mathrm{ax},\mathrm{co},\mathrm{sa}\}$ denotes the axial, coronal, and sagittal orthogonal views. $A_m(\cdot)$ denotes summation projection along the axis corresponding to view $m$, $\epsilon$ is a stability term, $E_{\mathrm{cnn}}(\cdot)$ denotes the shared 2D CNN, and $\beta_k^m$ denotes the view attention weight. Thus, TME converts each 3D local candidate region into a tri-view micro representation.

\subsection{Macro-Micro MIL Fusion Module}

With each ROI encoded by TME as a micro feature $f_k$, adaptive aggregation is required for patient-level prediction, as ROI-level contributions are not explicitly annotated. The \textbf{Macro-Micro MIL Fusion (MMF) module} treats multiple ROIs as a patient-level bag and models their relative importance through MIL attention. To maintain consistency between local features and full-volume MRI context, the macro feature generated by MRT module is also introduced into the final classification process, as illustrated in Fig.~\ref{fig:workflow}(d). For the $k$-th ROI, the center heatmap response score $s_k$ is first normalized within the same patient and incorporated into MIL attention. Specifically, the response-aware attention score is defined as
\begin{equation}
\alpha_k = \phi(f_k) + \lambda_s \cdot \mathrm{Norm}(s_k),
\end{equation}
where $\phi(\cdot)$ denotes a learnable feature-to-attention projection, and $\mathrm{Norm}(\cdot)$ denotes patient-wise score normalization. The final ROI attention weight is computed by
\begin{equation}
a_k = \frac{\exp(\alpha_k)}{\sum_{j=1}^{K}\exp(\alpha_j)}.
\end{equation}
\noindent where $a_k$ denotes the MIL attention weight of the $k$-th ROI, $K$ denotes the number of selected ROIs, and $\lambda_s=0.25$ by default. This design allows the center heatmap to contribute not only to ROI mining but also to patient-level feature aggregation. The patient-level micro feature, macro-feature projection, and fused feature are formulated as:

\begin{equation}
\begin{aligned}
f_{\mathrm{mic}} &= \sum_{k=1}^{K} a_k f_k, \\
\tilde{f}_{\mathrm{mac}} &= \phi(f_{\mathrm{mac}}), \\
f_{\mathrm{fus}} &= \operatorname{Concat}(\tilde{f}_{\mathrm{mac}}, f_{\mathrm{mic}}).
\end{aligned}
\end{equation}
\noindent where $f_{\mathrm{mic}}$ denotes the patient-level micro feature aggregated from all ROI features, $\tilde{f}_{\mathrm{mac}}$ denotes the projected macro feature, and $f_{\mathrm{fus}}$ denotes the fused macro-micro patient-level representation. $\phi(\cdot)$ denotes the macro-feature projection layer, and $\operatorname{Concat}(\cdot)$ denotes feature concatenation. This representation is fed into an MLP classifier, followed by Softmax activation to obtain patient-level prediction probabilities:
\begin{equation}
\begin{aligned}
\mathbf{p} &= \operatorname{Softmax}(\operatorname{MLP}(f_{\mathrm{fus}})), \\
\hat{y} &= p_1, \quad \hat{y}\in[0,1], \\
\tilde{y} &= \arg\max_{c\in\{0,1\}} p_c .
\end{aligned}
\end{equation}
\noindent where $\hat{y}$ denotes the prediction confidence for the high-density class, $\tilde{y}$ denotes the discrete prediction label, and $p_c$ denotes the predicted probability of class $c$. Thus, MMF module fuses local micro features with global macro semantics at the patient level and produces the final prediction. Through this design, CHM-Net forms a coherent patient-level representation that integrates localized high-density features and whole-volume MRI context within a unified MRI-MDS framework.

\subsection{Optimization Strategy}

Center heatmap learning, local ROI encoding, and patient-level classification involve different optimization objectives, making direct joint optimization prone to unstable gradient interference. To mitigate this issue, CHM-Net adopts a \textbf{Density-guided Two-stage Optimization (DTO) strategy}. First, MRT module is pretrained in an unsupervised manner to obtain a stable macro-level center heatmap and global representation. The macro-micro classification branch is then trained with patient-level labels, enabling ROI mining, micro-feature encoding, and patient-level prediction to be progressively connected. The training procedure is shown in Fig.~\ref{fig:workflow}(e). 

\subsubsection{Unsupervised MRT module Pretraining}
\label{sec:mdt_pretraining}
Stable macro-level pretraining supports subsequent ROI mining and patient-level stratification. Specifically, each input sample $x_i$ is randomly augmented into two views, $x_i^1$ and $x_i^2$. The two views are passed through MRT module and a projection head to obtain representations $z_i^1$ and $z_i^2$, respectively. This process adopts the NT-Xent contrastive loss \cite{chen2020simple}:
\begin{equation}
\begin{aligned}
\mathcal{L}_{\mathrm{con}}
&=
\frac{1}{2B}\sum_{i=1}^{B}
\left[
\ell(z_i^1,z_i^2)+\ell(z_i^2,z_i^1)
\right], \\
\ell(z_i^a,z_i^b)
&=
-\log
\frac{
\exp(\operatorname{sim}(z_i^a,z_i^b)/\tau)
}{
\sum_{j=1}^{B}
\exp(\operatorname{sim}(z_i^a,z_j^b)/\tau)
}.
\end{aligned}
\end{equation}
\noindent where $B$ denotes the batch size, $\operatorname{sim}(\cdot,\cdot)$ denotes cosine similarity, and $\tau$ denotes the temperature coefficient, with $\tau=0.1$ by default. By minimizing $\mathcal{L}_{\mathrm{con}}$, MRT module learns stable macro-level spatial responses without manual masks, thereby providing priors for subsequent ROI mining.

\subsubsection{Patient-level MIL Classification Optimization}
\label{sec:mil_optimization}
To translate the unsupervised heatmap prior into stratification performance, patient-level labels supervise CHM-Net classification learning. The classification loss is formulated as cross-entropy with label smoothing, combined with center heatmap sparsity and smoothness constraints:
\begin{equation}
\begin{aligned}
\mathcal{L}
&=
\mathcal{L}_{\mathrm{cls}}
+
\lambda_{\mathrm{sp}}\mathcal{L}_{\mathrm{sp}}
+
\lambda_{\mathrm{sm}}\mathcal{L}_{\mathrm{sm}}, \\
\mathcal{L}_{\mathrm{cls}}
&=
\operatorname{CE}_{\mathrm{ls}}(\mathbf{p},y), \\
\mathcal{L}_{\mathrm{sp}}
&=
\frac{1}{|\Omega|}
\sum_{v \in \Omega} h(v), \\
\mathcal{L}_{\mathrm{sm}}
&=
\frac{
\|\nabla_z h\|_1+
\|\nabla_y h\|_1+
\|\nabla_x h\|_1
}{3}.
\end{aligned}
\end{equation}
\noindent where $\mathcal{L}_{\mathrm{sp}}$ constrains the sparsity of the center heatmap, while $\mathcal{L}_{\mathrm{sm}}$ encourages its spatial continuity. $\operatorname{CE}_{\mathrm{ls}}(\cdot)$ denotes label-smoothing cross-entropy, with the smoothing coefficient set to $0.10$ by default. $\lambda_{\mathrm{sp}}$ and $\lambda_{\mathrm{sm}}$ are the loss weights, set to $0.01$ and $0.05$ by default, respectively. $|\Omega|$ denotes the total number of voxels, and $\nabla_z$, $\nabla_y$, and $\nabla_x$ denote first-order differences along the three spatial directions. Overall, DTO decouples center heatmap pretraining from patient-level classification optimization, thereby improving the training stability of ROI-level micro encoding and final classification.

\section{Experiments}

\subsection{Datasets}
\subsubsection{GBNPC 2026 Dataset}

\begin{figure}[!htbp]
    \centering
    \includegraphics[width=\linewidth]{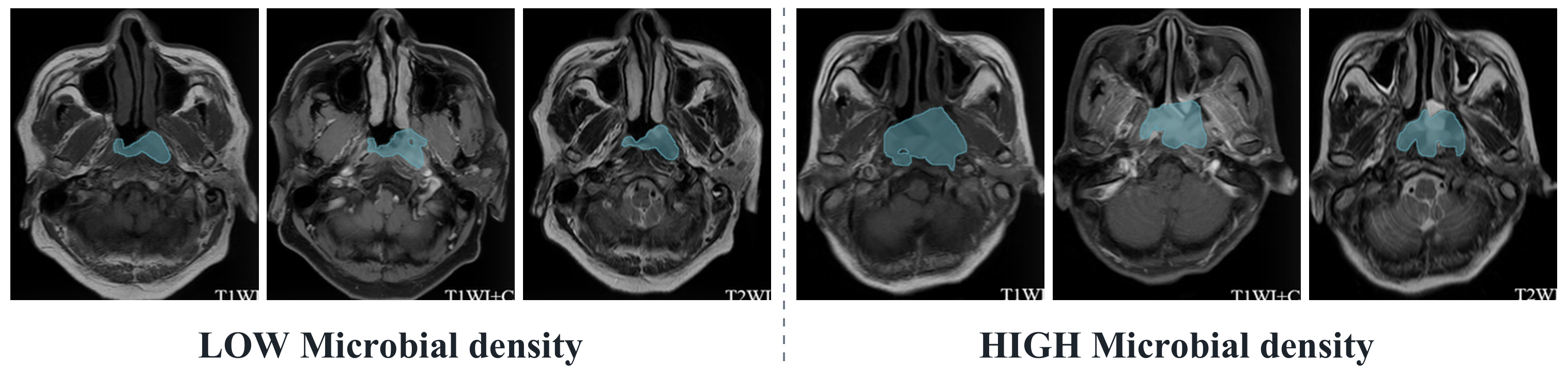}
    \caption{Representative visualization of GBNPC 2026 dataset. Blue denotes the ground truth.}
    \label{fig:gbnpc-visualization}
\end{figure}

The GBNPC 2026 (Germ Burden in Nasopharyngeal Carcinoma 2026) dataset is established to support non-invasive prediction of intratumoral microbial burden from multimodal MRI. This retrospective cohort includes 183 treatment-naive patients with pathologically confirmed nasopharyngeal carcinoma treated between December 2013 and January 2026. Cases with poor MRI quality, concomitant malignancies, or severe autoimmune diseases are excluded. The dataset consists of pretreatment multimodal MRI scans and intratumoral microbial burden labels. The MRI scans include T1-weighted, T2-weighted, and contrast-enhanced T1-weighted sequences, while labels are obtained by fluorescence in situ hybridization using the universal bacterial EUB338 probe. Representative MRI samples from GBNPC 2026 are shown in Fig.~\ref{fig:gbnpc-visualization}. 
\begin{table*}[!tbp]
\centering
\caption{Performance comparison on GBNPC 2026 under standard five-fold cross-validation. Highlighted values denote the best results, and underlined values denote the second-best results.}
\label{tab:private_dataset_results}
\small
\setlength{\tabcolsep}{5.0pt}
\renewcommand{\arraystretch}{1.2}
\resizebox{0.8\textwidth}{!}{%
\begin{tabular}{lccccc}
\toprule[1.2pt]
\textbf{Methods} & \textbf{ACC (\%)}$\uparrow$ & \textbf{AUC (\%)}$\uparrow$ & \textbf{F1 (\%)}$\uparrow$ & \textbf{Sens. (\%)}$\uparrow$ & \textbf{Spec. (\%)}$\uparrow$ \\
\midrule[0.9pt]
ResNet-50 + 2.5D\cite{he2016deep} & $49.19 \pm 0.70$ & $68.01 \pm 4.80$ & $34.44 \pm 3.20$ & $57.78 \pm 52.90$ & $42.11 \pm 53.00$ \\
ResNet-50 + 3D\cite{hara2018can} & $\underline{55.69 \pm 6.30}$ & $\underline{69.66 \pm 4.80}$ & $45.31 \pm 13.60$ & \cellcolor{blue!15}\textbf{$80.00 \pm 29.80$} & $30.58 \pm 42.10$ \\
ResNet-50 + ACS\cite{yang2021reinventing} & $54.08 \pm 5.30$ & $64.75 \pm 3.80$ & $44.81 \pm 11.20$ & $53.04 \pm 43.40$ & $55.96 \pm 46.40$ \\
auto-sklearn\cite{feurer2015efficient} & $52.96 \pm 6.02$ & $57.23 \pm 7.05$ & $49.38 \pm 10.67$ & $38.30 \pm 22.06$ & $67.49 \pm 20.50$ \\
X3D\cite{feichtenhofer2020x3d} & $48.06 \pm 3.75$ & $51.19 \pm 7.98$ & $48.08 \pm 19.35$ & $58.89 \pm 39.21$ & $36.43 \pm 34.54$ \\
AMSNet\cite{wu2022attention} & $49.74 \pm 3.31$ & $52.96 \pm 9.05$ & $49.00 \pm 19.91$ & $59.88 \pm 37.75$ & $40.47 \pm 39.21$ \\
M3T\cite{jang2022m3t} & $49.73 \pm 1.13$ & $50.21 \pm 9.29$ & $39.76 \pm 36.30$ & $60.00 \pm 54.77$ & $40.00 \pm 54.77$ \\
3DCT-ICH\cite{xiong2024multimodality} & $50.81 \pm 3.39$ & $49.90 \pm 8.74$ & $36.77 \pm 33.95$ & $49.47 \pm 50.01$ & $53.33 \pm 50.55$ \\
XFMamba\cite{zheng2025xfmamba} & $53.00 \pm 1.12$ & $57.44 \pm 2.27$ & $\underline{55.45 \pm 8.53}$ & $61.40 \pm 17.37$ & $44.27 \pm 17.76$ \\
SwinTransformer-3D\cite{liu2022video} & $50.81 \pm 0.74$ & $54.99 \pm 7.52$ & $13.57 \pm 30.35$ & $20.00 \pm 44.72$ & \cellcolor{blue!15}\textbf{$80.00 \pm 44.72$} \\
Med3D\cite{chen2019med3d} & $49.16 \pm 3.41$ & $53.67 \pm 7.24$ & $24.22 \pm 28.22$ & $27.78 \pm 41.94$ & $\underline{71.29 \pm 37.53}$ \\
MedVit-3D\cite{manzari2023medvit} & $49.20 \pm 6.53$ & $44.92 \pm 8.31$ & $38.85 \pm 26.33$ & $45.56 \pm 39.17$ & $52.51 \pm 43.91$ \\
\midrule[0.9pt]
\textbf{CHM-Net} & \cellcolor{blue!15}\textbf{$67.75 \pm 2.46$} & \cellcolor{blue!15}\textbf{$69.69 \pm 1.19$} & \cellcolor{blue!15}\textbf{$66.47 \pm 4.19$} & $\underline{64.74 \pm 6.80}$ & $70.64 \pm 2.95$ \\
\bottomrule[1.2pt]
\end{tabular}%
}
\end{table*}

\begin{table*}[!tbp]
\centering
\caption{Comparison results on two public datasets. Highlighting and underlining follow the notation in Table~\ref{tab:private_dataset_results}.}
\label{tab:integrated_results}
\scriptsize
\setlength{\tabcolsep}{2.8pt}
\renewcommand{\arraystretch}{1.2}
\resizebox{0.8\textwidth}{!}{%
\begin{tabular}{l|ccccc|ccccc}
\toprule[1.2pt]
\multirow{2}{*}{\textbf{Methods}}
& \multicolumn{5}{c|}{\textbf{NoduleMNIST3D (\%)}}
& \multicolumn{5}{c}{\textbf{AdrenalMNIST3D (\%)}} \\
\cmidrule(lr){2-6}\cmidrule(lr){7-11}
& ACC$\uparrow$ & AUC$\uparrow$ & F1$\uparrow$ & Sens.$\uparrow$ & Spec.$\uparrow$
& ACC$\uparrow$ & AUC$\uparrow$ & F1$\uparrow$ & Sens.$\uparrow$ & Spec.$\uparrow$ \\
\midrule[0.9pt]
ResNet-50 + 2.5D\cite{he2016deep} & $84.84$ & $83.53$ & $58.66$ & $52.08$ & \cellcolor{blue!15}\textbf{$93.36$} & $46.98$ & $\underline{81.91}$ & $47.53$ & \cellcolor{blue!15}\textbf{$97.10$} & $31.88$ \\
ResNet-50 + 3D\cite{hara2018can} & $84.73$ & $87.49$ & $61.75$ & $59.90$ & $91.19$ & $80.54$ & $72.01$ & \cellcolor{blue!15}\textbf{$81.68$} & $82.61$ & $79.91$ \\
ResNet-50 + ACS\cite{yang2021reinventing} & $84.09$ & $88.55$ & $63.75$ & $67.71$ & $88.35$ & $76.51$ & $74.43$ & $\underline{78.17}$ & $\underline{84.06}$ & $74.24$ \\
auto-sklearn\cite{feurer2015efficient} & $87.42$ & \cellcolor{blue!15}\textbf{$91.40$} & $68.39$ & $66.15$ & $\underline{92.95}$ & $78.86$ & $81.50$ & $59.15$ & $20.29$ & $96.51$ \\
X3D\cite{feichtenhofer2020x3d} & $83.23$ & $87.86$ & $83.33$ & $83.87$ & $82.58$ & $81.88$ & $63.91$ & $43.75$ & $30.43$ & $97.38$ \\
AMSNet\cite{wu2022attention} & $84.52$ & $88.47$ & $84.62$ & $85.16$ & $83.87$ & $\underline{86.58}$ & $74.05$ & $63.64$ & $50.72$ & $97.38$ \\
M3T\cite{jang2022m3t} & $84.84$ & $88.93$ & $84.69$ & $83.87$ & $85.81$ & $83.22$ & $65.29$ & $46.81$ & $31.89$ & $\underline{98.69}$ \\
3DCT-ICH\cite{xiong2024multimodality} & $86.13$ & $89.89$ & $86.35$ & $87.74$ & $84.52$ & $84.90$ & $67.90$ & $52.63$ & $36.23$ & \cellcolor{blue!15}\textbf{$99.56$} \\
XFMamba\cite{zheng2025xfmamba} & $86.45$ & $90.27$ & $86.27$ & $85.16$ & $87.74$ & $82.89$ & $67.60$ & $51.43$ & $39.13$ & $96.07$ \\
SwinTransformer-3D\cite{liu2022video} & $86.77$ & $90.76$ & $86.98$ & $\underline{88.39}$ & $85.16$ & $82.21$ & $65.64$ & $47.52$ & $34.78$ & $96.51$ \\
Med3D\cite{chen2019med3d} & $87.10$ & $90.36$ & $87.01$ & $86.45$ & $87.74$ & $84.56$ & $72.24$ & $59.65$ & $49.28$ & $95.20$ \\
MedVit-3D\cite{manzari2023medvit} & $\underline{87.74}$ & $91.10$ & $\underline{87.66}$ & $87.10$ & $88.39$ & $79.53$ & $63.90$ & $44.04$ & $34.78$ & $93.01$ \\
\midrule[0.9pt]
\textbf{CHM-Net} & \cellcolor{blue!15}\textbf{$88.39$} & $\underline{91.38}$ & \cellcolor{blue!15}\textbf{$88.46$} & \cellcolor{blue!15}\textbf{$89.03$} & $87.74$ & \cellcolor{blue!15}\textbf{$87.58$} & \cellcolor{blue!15}\textbf{$87.74$} & $70.87$ & $65.22$ & $94.32$ \\
\bottomrule[1.2pt]
\end{tabular}%
}
\end{table*}

\subsubsection{Auxiliary Validation Datasets}

To further evaluate model generalization, auxiliary validation was performed on two 3D medical image datasets, NoduleMNIST3D and AdrenalMNIST3D~\cite{yang2023medmnist}. Specifically, NoduleMNIST3D contains 1,633 lung nodule CT samples for benign-malignant classification of pulmonary nodules, while AdrenalMNIST3D contains 1,584 adrenal CT samples for binary classification of adrenal lesions. Both datasets provide standardized 3D medical images with binary labels, and are therefore suitable for evaluating the cross-domain adaptability of the proposed macro-micro modeling framework across different organs and lesion types.

\subsection{Implementation Details}

The proposed framework was implemented using PyTorch, and all experiments were conducted on four NVIDIA GeForce RTX 2080 Ti GPUs. The training procedure followed the DTO strategy. First, the MRT module was pretrained in an unsupervised manner for 100 epochs with a batch size of 4 and a learning rate of $1\times10^{-4}$. Subsequently, in the patient-level classification stage, the model was trained for 150 epochs with AdamW optimizer, with $\beta_1=0.9$, $\beta_2=0.999$, and a weight decay of $1\times10^{-4}$. In this stage, the batch size was also set to 4, the initial learning rate was set to $5\times10^{-5}$, and a cosine annealing schedule was adopted for learning rate decay. The loss weights were set to $\lambda_{\mathrm{sp}}=0.01$ and $\lambda_{\mathrm{sm}}=0.05$.

\subsection{Compared Methods}

To systematically evaluate the effectiveness of \mbox{CHM-Net}, several standard medical image classification methods were selected for comparison. Specifically, the CNN baselines comprised ResNet-50+2.5D~\cite{he2016deep}, ResNet-50+3D~\cite{hara2018can}, ResNet-50+ACS~\cite{yang2021reinventing}, and X3D~\cite{feichtenhofer2020x3d}, covering different CNN adaptations to volumetric inputs and efficient 3D convolution. auto-sklearn~\cite{feurer2015efficient} was included as an automated machine learning baseline. For medical volumetric representation learning, AMSNet~\cite{wu2022attention}, 3DCT-ICH~\cite{xiong2024multimodality}, and Med3D~\cite{chen2019med3d} were selected to represent multi-scale modeling, multimodal 3D modeling, and medical 3D pretraining. Transformer-based and Mamba-based architectures, including M3T~\cite{jang2022m3t}, SwinTransformer-3D~\cite{liu2022video}, MedViT-3D~\cite{manzari2023medvit}, and XFMamba~\cite{zheng2025xfmamba}, cover multi-view representation, window self-attention, convolution-Transformer hybrid modeling, and long-range dependency modeling. These comparisons establish a rigorous benchmark for CHM-Net.

\begin{figure*}[!tbp]
\centering
\includegraphics[width=\textwidth]{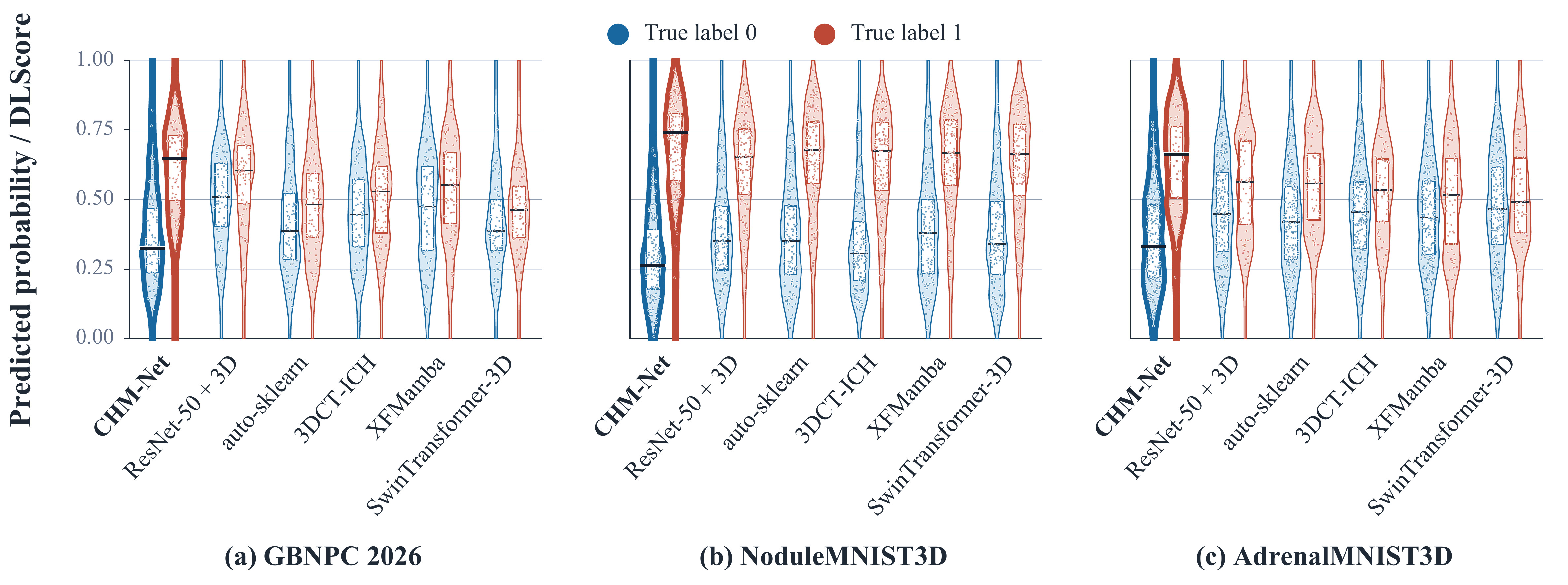}
\caption{Prediction-score violin plots on GBNPC 2026, NoduleMNIST3D, and AdrenalMNIST3D.}
\label{fig:violin_results}
\end{figure*}

\begin{figure}[!htbp]

\noindent
\begin{minipage}{\linewidth}
\includegraphics[width=\linewidth]{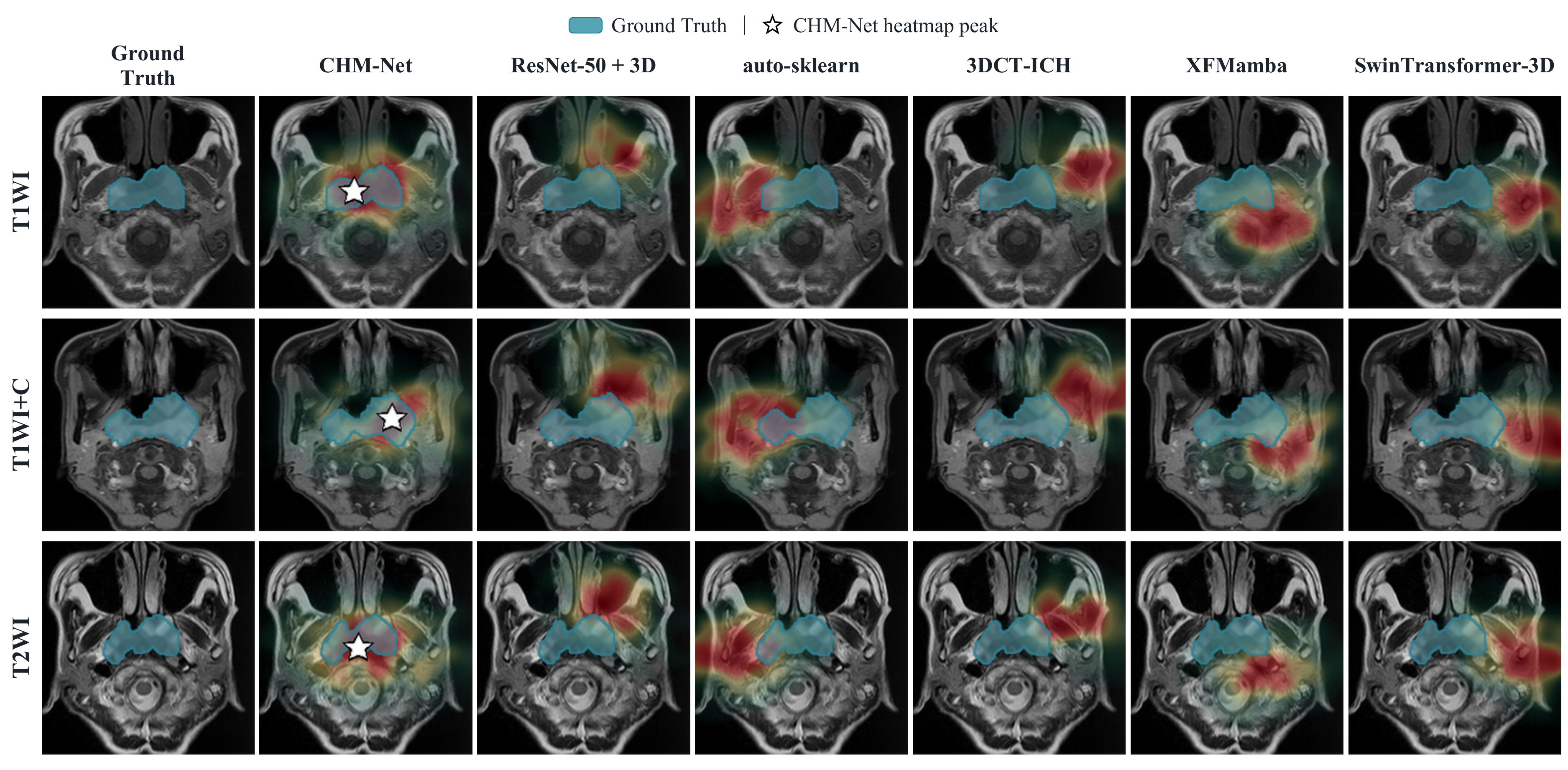}
\vspace{-2mm}
\centering{\footnotesize (a) GBNPC 2026}
\end{minipage}

\vspace{2mm}

\noindent
\begin{minipage}{\linewidth}
\includegraphics[width=\linewidth]{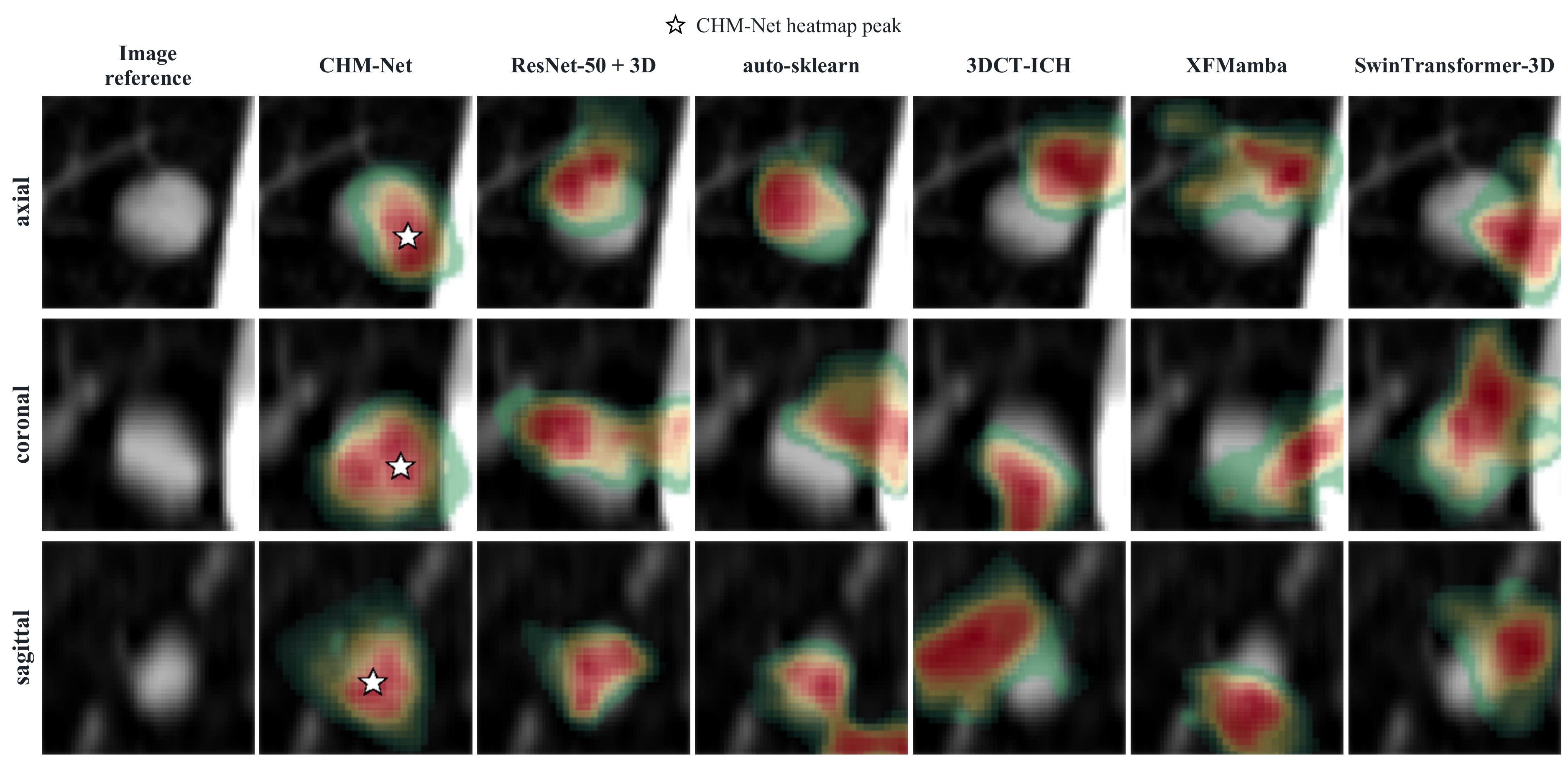}
\vspace{-2mm}
\centering{\footnotesize (b) NoduleMNIST3D}
\end{minipage}

\vspace{2mm}

\noindent
\begin{minipage}{\linewidth}
\includegraphics[width=\linewidth]{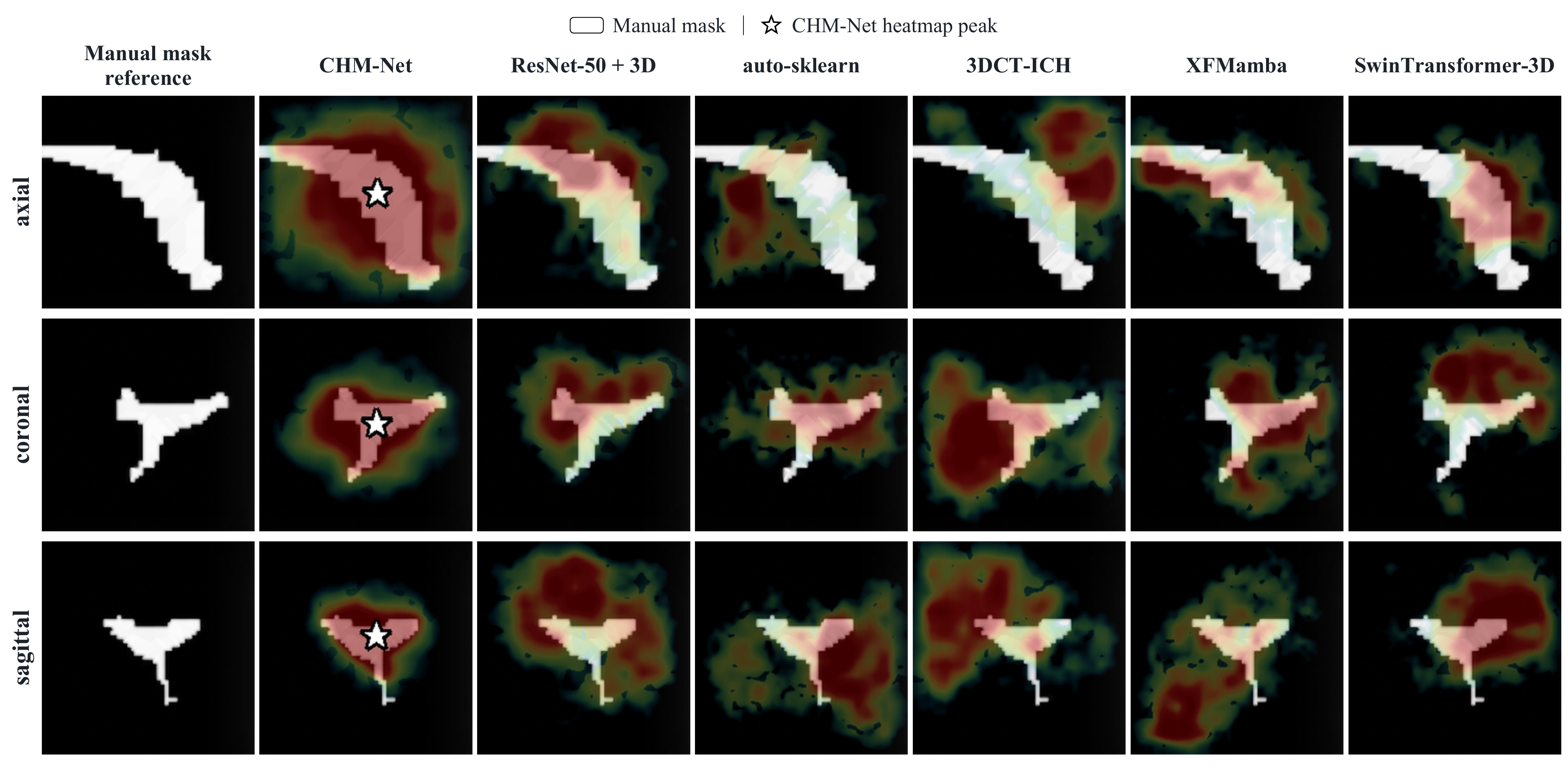}
\vspace{-2mm}
\centering{\footnotesize (c) AdrenalMNIST3D}
\end{minipage}

\caption{Class activation heatmaps on GBNPC 2026, NoduleMNIST3D, and AdrenalMNIST3D.}
\label{fig:heatmap_results}
\end{figure}

\subsection{Results and Discussion}


Table~\ref{tab:private_dataset_results} shows that CHM-Net achieves the best overall performance on GBNPC 2026 under five-fold cross-validation, with an AUC, ACC, and F1-score of 69.69\%, 67.75\%, and 66.47\%, respectively. Compared with the strongest competing results, CHM-Net improves ACC by 12.06 percentage points, AUC by 0.03 percentage points, and F1-score by 11.02 percentage points, yielding relative gains of 21.65\%, 0.04\%, and 19.87\%, respectively. Although ResNet-50+3D obtains a comparable AUC among CNN baselines, its ACC and F1-score are much lower. Transformer-based and Mamba-based models also fail to consistently produce robust patient-level discrimination. The maximum standard deviation across all metrics is 6.80 percentage points, indicating stable performance. These results suggest that CHM-Net learns a more discriminative macro-micro representation by coupling center-heatmap-guided ROI mining with patient-level feature fusion.

The auxiliary validation results further support the robustness of CHM-Net. As shown in Table~\ref{tab:integrated_results}, on NoduleMNIST3D, CHM-Net achieves the best result in ACC, F1-score, and Sensitivity, with improvements of 0.65, 0.80, and 0.64 percentage points over the second-best methods, corresponding to relative gains of 0.74\%, 0.91\%, and 0.72\%, respectively. On AdrenalMNIST3D, CHM-Net achieves the best AUC and ACC, reaching 87.74\% and 87.58\%, respectively. Compared with the second-best results, the improvements are 5.83 percentage points in AUC and 1.00 percentage point in ACC, corresponding to relative gains of 7.12\% and 1.15\%.\begin{table}[!htbp]
\centering
\scriptsize
\setlength{\tabcolsep}{2.0pt}
\renewcommand{\arraystretch}{1.15}
\caption{Ablation study under five-fold cross-validation. M1: detector-guided ROI mining; M2: ROI heatmap gating; M3: tri-planar encoder; M4: macro-micro fusion.}
\label{tab:ablation_study}
\begin{tabular}{lccccc}
\toprule
Config. & ACC$\uparrow$ & AUC$\uparrow$ & F1$\uparrow$ & Sens.$\uparrow$ & Spec.$\uparrow$ \\
\midrule
Base & 51.88$\pm$8.90 & 52.00$\pm$0.88 & 41.92$\pm$18.72 & 39.59$\pm$24.70 & 63.74$\pm$25.68 \\
+M1 & 57.39$\pm$0.86 & 59.30$\pm$6.24 & 46.82$\pm$6.03 & 38.42$\pm$8.47 & 76.02$\pm$7.55 \\
+M1+M2 & 59.01$\pm$0.62 & 60.36$\pm$1.57 & 49.51$\pm$3.18 & 40.64$\pm$4.68 & \cellcolor{blue!12}77.08$\pm$5.05 \\
+M1+M2+M3 & 63.93$\pm$1.10 & 63.84$\pm$1.68 & 59.68$\pm$2.56 & 53.80$\pm$3.60 & 73.86$\pm$2.91 \\
Full & \cellcolor{blue!12}\textbf{66.11$\pm$1.66} & \cellcolor{blue!12}\textbf{69.04$\pm$1.59} & \cellcolor{blue!12}\textbf{65.09$\pm$2.63} & \cellcolor{blue!12}\textbf{63.74$\pm$4.82} & 68.48$\pm$2.27 \\
\bottomrule
\end{tabular}
\end{table}

\begin{figure}[!htbp]
    \centering
    \includegraphics[width=\linewidth]{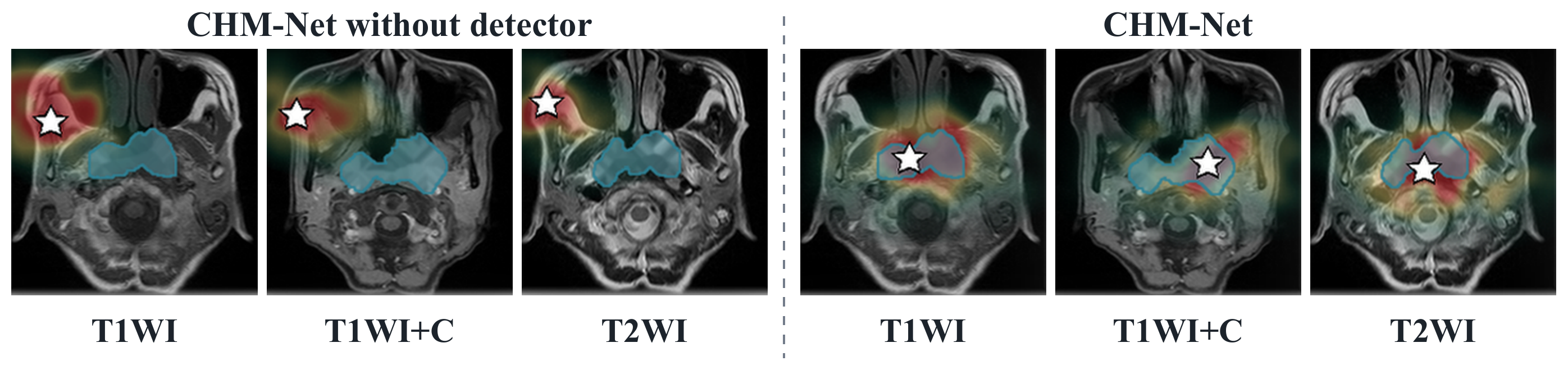}
    \caption{Detector ablation results on GBNPC 2026. Blue denotes the ground truth, \ding{73} indicates detected centers, and red denotes heatmap responses.}
    \label{fig:private-heatmap-detector-ablation}
\end{figure}
These findings indicate the generalizability of the proposed macro-micro modeling strategy to other volumetric medical image classification tasks.

The prediction score distributions in Fig.~\ref{fig:violin_results} further support the discriminative capacity of CHM-Net, showing clearer score separation and more reliable patient-level confidence.

\subsection{Qualitative Visualizations}


Fig.~\ref{fig:heatmap_results} visualizes attention heatmaps to analyze CHM-Net’s decision rationale. Since lesion morphology may relate to microbial density, an effective model should focus on lesion-related regions. Without the center-heatmap constraint, attention tends to drift toward non-discriminative areas, leading to scattered activations. In contrast, CHM-Net generates more concentrated responses around lesion regions, indicating its ability to capture discriminative MRI-MDS evidence. Similar patterns are also observed on the two auxiliary datasets.

\subsection{Ablation Study}

Table~\ref{tab:ablation_study} presents the stepwise ablation results of CHM-Net. The baseline achieves only 51.88\% ACC and 52.00\% AUC, indicating that fixed ROI selection and simple aggregation provide limited discriminative capability. As detector-guided ROI mining (M1), ROI heatmap gating (M2), tri-planar encoding (M3), and attention-based MIL aggregation with macro-micro fusion (M4) are progressively incorporated, performance improves consistently. The full CHM-Net achieves the best results, with 66.11\% ACC, 69.04\% AUC, and 65.09\% F1-score, demonstrating the joint effectiveness of the proposed components. Fig.~\ref{fig:private-heatmap-detector-ablation} further illustrates that the proposed design guides attention to key imaging regions.


\section{Conclusion}

This work introduces MRI-based microbial density stratification (MRI-MDS) as a novel non-invasive predictive task, aiming to infer postoperative microbial density status from preoperative multimodal MRI. To address MRI-MDS, CHM-Net is introduced as a center heatmap-driven macro-micro modeling framework. Specifically, CHM-Net derives a center heatmap from global MRI representations as a response prior, thereby highlighting subtle lesion-related regions in MRI images. The localized responses are then distilled into patient-level macro-micro evidence by integrating fine-grained regional patterns with global MRI semantics. To support baseline evaluation, this work constructs GBNPC 2026, a multimodal MRI dataset of nasopharyngeal carcinoma patients. Extensive experiments demonstrate CHM-Net outperforms multiple representative baselines on GBNPC 2026 and exhibits robustness on NoduleMNIST3D and AdrenalMNIST3D. Collectively, these results establish MRI-MDS as a feasible and clinically meaningful non-invasive task, while positioning CHM-Net as an effective, interpretable, and extensible baseline for imaging-based microbial density assessment.



\FloatBarrier

\vspace{12pt}


\end{document}